\documentstyle[amssymb,graphicx,epsfig,aps,twocolumn]{revtex}
\begin{document}

\twocolumn[\hsize\textwidth\columnwidth\hsize\csname 
@twocolumnfalse\endcsname

\title{Stripes: Why hole rich lines are antiphase domain walls?}
\author{Oron Zachar}
\address{ICTP, 11 strada Costiera, Trieste 34100, Italia}
\date{\today}
\maketitle

\widetext
\begin{abstract}

For stripes of hole rich lines in doped antiferromagnets, we investigate the 
competition between anti-phase and in-phase domain wall ground state 
configurations. We argue that a phase transition must occure as a function 
of the electron/hole filling fraction of the domain wall. Due to {\em transverse} 
kinetic hole fluctuations, empty domain walls are always anti-phase. 
At arbitrary electron filling fraction ($\delta $) of the domain wall 
(and in particular for $\delta \approx 1/4$ as in LaNdSrCuO), it is essential 
to account also for the transverse magnetic interactions of the electrons and their 
mobility {\em along} the domain wall.

We find that the transition from anti-phase to in-phase stripe domain wall occurs 
at a critical filling fraction $0.28<\delta _{c}<0.30$, for any value of 
$\frac{J}{t}<\frac{1}{3}$. We further use our model to estimate the spin-wave 
velocity in a stripe system. Finally, we relate the results of 
our microscopic model to previous Landau theory approach to stripes.

\smallskip
\end{abstract}

 ]

\narrowtext

\section{Introduction}

An anti-phase domain wall in stripes is the state where the local
antiferromagnetic (AFM) spin order parameter undergo a $\pi $ phase shift
across a hole rich line. Such periodic stripe structures were experimentally
found in doped Copper oxides and Nickel oxides\cite{Tranquada-reviews}.
Historically, this well accepted feature was first considered to be a
natural outcome of mean-field theory Fermi surface instability\cite
{zaanen,Note1}. Yet, no similar rigorous microscopic explanation was given
within the frustrated phase separation picture\cite
{FrustratedPhaseSeparation} which is currently regarded as underlying the
microscopic origin of stripes in the relevant materials\cite{StripesLandau}.
One of our goals here is to close this gap in the theory.

In contrast with common folklore, we show that hole rich lines are not
necessarily anti-phase domain walls of AFM spin domains. First, on simple
general grounds, we argue that there must be a phase transition from
anti-phase to in-phase domain wall configuration as a function of increased
electron filling fraction $\delta $ of the domain wall. We then proceed to
construct microscopic t-J models of the local electronic dynamics which
determines the resulting spin order across a hole line.

It is frequently argued theoretically\cite
{FrustratedPhaseSeparation,StripesLandau} and exemplified experimentally\cite
{Tranquada-reviews} that the charge segregation into hole rich stripe lines
is prior to the establishment antiphase spin domains. Therefore, the
microscopic mechanism by which the hole lines enslave the spin order should
be distinguished and considered separately.

We develope a qualitative and quantitative microscopic understanding of the
domain wall magnetic order. Our analysis accounts for the electronic
dynamics both transverse and along a pre-established hole rich line between
two AFM domains. Thus we focus solely on the mechanism which give rise to
the spin antiphase domain wall feature, by examining the competition between
anti-phase and in-phase configurations at given electron filling fraction $%
\delta $ of the hole rich line.

In the stripes literature it is more common to describe the domain wall
filling in terms of the number of holes (below half-filling) per site along
the domain wall; 
\begin{equation}
n^{h}=1-2\delta .
\end{equation}
Experimentally\cite{Tranquada-reviews}, anti-phase domain wall stripes in
Nickel-Oxides have $n^{h}\approx 1$ (corresponding to one hole per site
along the domain wall, or equivalently an electron empty domain wall, $%
\delta =0$), while anti-phase domain wall stripes in Copper-Oxides have $%
n^{h}\approx \frac{1}{2}$ (corresponding to one hole per two sites along the
domain wall, or electron 1/4 filled domain wall, $\delta \approx 0.25$).
These domain wall filling fractions remain roughly constant over a wide
range of dopings in the respective materials.

We find that the transition from anti-phase to in-phase stripe domain wall
occurs at critical filling fraction 
\begin{equation}
0.28<\delta _{c}<0.30
\end{equation}
depending on the value of $\left( \frac{J}{t}\right) $. In other words, for
example, we predict that stripe domain walls with a hole density of one hole
per three sites are always in-phase and not anti-phase. Appropriate
numerical simulations can be constructed to get exact numbers beyond the
limits of our analytical approximations. Yet, since our analysis indicates
that the sensitivity to $\left( \frac{J}{t}\right) $ is rather weak, we do
not expect the exact numerical results to deviate much from our predictions.

We further apply our model to evaluate the of spin-wave velocity in a stripe
systems, and compare with experiments. In addition, we relate the competing
interactions in our microscopic model to previous Landau theory approach to
stripes order\cite{StripesLandau,Leonid99-StripesLandau}. Thus, we advance
towards a coherent microscopic and phenomenological understanding of the
stripes structure within the frustrated phase-separation picture.

\subsection{Overview of the model construction, analysis and main results}

Since the paper is quite long, we here supply the reader with an overview of
the gist of our model development and main results.

The general intuitive argument for the domain wall transition is the
following: In one extreme case, $n^{h}=1$, where there is a hole on each
site along the domain wall (i.e., it is empty of electrons, $\delta =0$, as
in Nickel-Oxides), it is clear that an anti-phase domain wall configuration
is favored by transverse hole fluctuations. Now, consider the state of
domain walls with increasing electron filling fraction. In the opposite
extreme case where the domain wall is half-filled with electrons (i.e., $%
n^{h}=0$, $\delta =\frac{1}{2}$), we should recover the undoped ordered AFM
state. Therefore, {\em there must be some intermediate critical electron
filling fraction }$\delta _{c}${\em \ of the domain wall at which there is a
transition from an anti-phase to in-phase local AFM spin configuration
across a hole rich line}.

The main quantitative objective of this paper is to determine the critical
domain wall filling fraction $\delta _{c}$ as a function of the t-J model
parameters. In the process, we develope an advanced qualitative and
quantitative understanding of the various competing local interactions in
the single stripe physics. Additional applications are discussed in
sections-V and VI.

Intuitively, increase of electron filling of the domain wall amounts to
increase of magnetic interactions across the domain wall until they are
strong enough to dominate over the charge fluctuation dynamics (due to
holes). The possibility of a transition from topological (anti-phase) to
non-topological (in-phase) stripes, due to increase of AFM interactions, was
first speculated by Neto \& Hone\cite{Neto-Hone,NetoHone-Unpub} (though, on
not quite rigorous grounds, it was somehow related to the spin correlation
length).

We find that the topological/non-topological nature of the stripe charge
wall can be completely determined by the local dynamics, which in turn is
determined by the electron filling fraction of the wall (assuming fixed
given t-J model parameters). We do not see a way by which interaction
between domain walls, the width of the spin domains or any similar long
length scale are significantly relevant.

In section-II, we list some preliminary assumptions of our model of the
stripe domain wall: (a) The effective electron dynamics is captured by a one
band t-J model for electrons hopping between Copper sites (i.e., in which
the Oxygen sites are integrated out). (b) It is assumed that the hole line,
its mean position, and hole density are already pre-established by some
higher energy processes (presumably phase separation and coulomb
frustration). We hence focus solely on determining the preferred spin
configuration across a pre-established hole line. (c) The spin order is
determined by comparing the energetics of anti-phase and in-phase domain
wall configurations (and not with the motion of dilute holes in a uniform
AFM as was done in most previous work\cite
{Nayak-stripes,Neto-StripeStability}).

In section-III, we introduce our microscopic model and start with the
consideration of only {\em transverse} interactions and dynamics, (i.e.,
ignoring the effects of dynamics along the domain wall). The kinetic energy
due to transverse hopping of holes favors an anti-phase domain wall
configuration, while magnetic interaction between electrons favors an
in-phase domain configuration (see figure-1 below). These competing
interactions determine the preferred local spin configuration. All
calculations are done to the first significant order in $\left( \frac{J}{t}%
\right) $. Our analysis surprisingly shows that if dynamics along the domain
wall is ignored then transition to an in-phase domain wall would have
occurred already at $n^{h}\leq 2/3$ (i.e., at a density of two holes per
three sites), in conflict with experimental observations of anti-phase
stripes in $\left( LaNd\right) SrCuO$ materials with $n^{h}\approx 1/2$.
Hence, it is {\em essential} to investigate the effect of kinetic dynamics 
{\em along} the domain wall, which we undertake in section-IV.

In section-IV, we model the kinetic dynamics of the electrons moving along a
stripe domain wall in an effective external magnetic mean field due to its
AFM environment. Along an in-phase domain wall there is an effective net
staggered external magnetic field, while along an anti-phase domain wall the
net external field is zero on each site. We analyze two extreme limits: (a)
Non-interacting electrons moving along the domain wall, and (b) Large $U\gg
t $ limit for the interaction of electrons along the domain wall. In both
cases, the essential result is that kinetic fluctuations along the domain
wall weaken the average magnetic interaction energy which favors an in-phase
domain wall, and thus extend the stability of an anti-phase domain wall
configuration to higher electron filling fractions (i.e., lower hole
densities).

In section-V, we relate the competing interactions in our microscopic model
to previous Landau theory approach to stripes order \cite
{StripesLandau,Leonid99-StripesLandau}.

In section-VI, we apply our model to evaluate the spin-wave velocity $%
v_{\perp }$ in a stripe state compared with $v_{0}$ in the parent AFM
material (where $v_{\perp }$ is velocity perpendicular to the stripes).

We argue that our results are quantitatively accurate well beyond the
seemingly rough approximations of our simple models. The heart of the matter
is the fact that our quantitative results are sensitive only to the
energetic {\em difference} between an anti-phase and in-phase domain wall
configuration. Processes which are neglected in our treatment (e.g., deeper
penetration of hole hopping into the AFM environment) have the same
contribution in both domain wall configurations and thus only very weakly
affect the energetic difference between them.

\section{Preliminary assumptions}

The stripes characteristics can vary to include both diagonal and vertical
stripes, which may be insulating or conducting. Therefore, the anti-phase
domain wall mechanism should be rather simple, robust, and not too sensitive
to the above mentioned variations.

\begin{enumerate}
\item  The most microscopic Hubbard type model of the CuO$_{2}$ or NiO$_{2}$
planes includes distinct Oxygen and Copper (or Nickel) orbitals bands. Yet,
as commonly argued, we assume that the effective dynamic is captured by a
one-band t-J model\cite{ZhangRice} (in which the oxygen sites degrees of
freedom have been integrated out) where electrons hop directly between
Copper lattice sites. In the context of stripes theory, the above assumption
is supported by the fact that correct domain wall configurations turned out
in numerical simulations\cite{Manousakis98-stripes} of one band t-J models.

\item  Experimental evidence\cite{Tranquada-reviews} and theoretical
considerations\cite{StripesLandau} suggest that stripe formation is commonly
charge driven, i.e., that periodic hole line stripes form first and enslave
the formation of the AFM spin domain. Therefore, for our purpose in this
paper we take the hole lines to be pre-formed.\ 

\item  Stripes were found in both Spin-$\frac{1}{2}$ and Spin-$1$ doped
antiferromagnets\cite{Tranquada-reviews}. \ A doped hole can thus correspond
to a spin-$0$ or a spin-$\frac{1}{2}$ site respectively. \ Yet, in both
cases the hole and its dynamics are carried on only within the $%
d_{x^{2}-y^{2}}$ orbital band. \ Indeed, the model which we construct and
analyze works equally well for both doped Spin-$\frac{1}{2}$ and Spin-$1$
antiferromagnets.\ We chose to present our analysis in terms of a doped Spin-%
$\frac{1}{2}$ AFM (i.e., corresponding to stripes in a CuO$_{2}$ plane).

\item  As with superexchange mechanism of antiferromagnetism in the undoped
parent system, we argue that the domain wall spin order is determined by
local interactions across the hole rich line. Hence, it is sufficient to
consider a single stripe segment in isolation (see figure-1).

\item  There's a non-trivial distinction between site-centered and
bond-centered domain walls\cite{TranquadaNiO-BondWall}, in the sense that
the spin alignment across an anti-phase domain wall is antiferromagnetic for
site-centered stripe and ferromagnetic for bond-centered stripe. The
presentation in this paper is conducted in terms of site-centered stripes,
and elsewhere\cite{Zachar-unpublished} we will show that the same principles
apply for bond-centered stripes as well.

\item  Probably the main quantitative approximation in our model is that we
treat the AFM regions between the hole lines as if they were
antiferromagnetically ordered. We neglect spin exchange fluctuations and the
quantum nature of the AFM correlations in the rather narrow ladder geometry
of the stripes. In other words, our quantitative results are rigorously
valid for an Ising model approximation of the AFM regions. Yet, in all of
our calculations we use a parameter $\varepsilon $ which is defined to be
the energy difference between parallel and anti-parallel near-neighbor spin
states (see equation (\ref{epsilon-define})). Only in the end we substitute $%
\varepsilon =J$ for intuitive concreteness. In principle, all the effects of
fluctuations etc. can be incorporated into a renormalized value of $%
\varepsilon $ without changing our results.

\item  Our quantitative results are sensitive only to the energetic {\em %
difference} between an anti-phase and in-phase domain wall configuration.
Therefore, many processes which are neglected in our treatment (e.g., deeper
penetration of hole hopping into the AFM environment, magnetic interaction
between electrons on the wall) have the same contribution in both domain
wall configurations and thus only very weakly affect the energetic
difference between them.
\end{enumerate}

\section{The effect of {\em transverse} interaction\label%
{Section - Transverse only}}

Our analysis proceed in two stages; First, in this section, we consider
solely the {\em transverse} fluctuations of a hole and magnetic interaction
of electrons in the stripe, while the holes/electrons configuration along
the domain wall is taken as static. In a second stage (section-IV), we
consider the implications of electron mobility {\em along} the stripe.

\subsection{Model of site-centered stripe}

The average hole line position is fixed by introducing a chemical potential
shift ($\mu $) on particular sites\cite{Neto-Hone,NetoHone-Unpub}. (In \
Figure-1, hole line $\mu $-shifted sites are represented by dark circles).\
In the ground state, holes are preferentially situated on the $\mu $ shifted
sites.\ Hence, the effective local chemical potential shift, $\mu $,
incorporates the net effect of the high energy dynamics (e.g., coulomb
frustrated phase separation \cite{FrustratedPhaseSeparation}) which lead to
the stripe formation. The magnitude of $\mu $ determines the stripe
stability, and can be associated with the stripe creation temperature for
the purpose of extracting an experimental estimate of it's magnitude. Thus,
it is assumed that $\mu <J<t$.

An antiferromagnetic spin exchange interaction $J$ exists between electron
spins on nearest neighbors sites. In order to fix the spin order, it is
enough to determine the relative orientation of the spins immediately to the
left and to the right of the hole line. Therefore, we include a {\sl %
transverse} hoping interaction, $t$, only between a domain wall site
(indicated by a dark circle in figure-1a) and its right and left nearest
neighbors in the AFM regions. As noted before, we argue that processes of
further penetration of a hole into the AFM regions are not significantly
affecting our results.\ Hopping interaction $t_{\parallel }$ {\sl along} the
domain wall will be considered in section-IV.

In the absence of kinetic motion along the domain wall, the Hamiltonian
describing a single site centered domain wall is 
\begin{eqnarray}
H &=&\sum\limits_{i}H_{i} \\
H_{i} &=&\mu n_{0,i}+J\frac{1}{2}\sum\limits_{\left\langle jj^{^{\prime
}}\right\rangle \left\langle ii^{^{\prime }}\right\rangle }\text{{\bf S}}%
_{j,i}\cdot \text{{\bf S}}_{j^{\prime },i^{\prime }}  \label{H_site} \\
&&-t\sum\limits_{i,s}\left[ \left( c_{1,i,s}^{\dagger
}c_{0,i,s}+c_{-1,i,s}^{\dagger }c_{0,i,s}\right) +{\rm h.c.}\right] 
\nonumber
\end{eqnarray}
where {\bf S}$_{j,i}$ is the spin of an electron at column $j$ and line $i$
(where $j=0$ is the domain wall position). $n_{0,i}$ is the occupation
number of a domain wall site at position $i$ along the wall. $%
c_{1,i,s}^{\dagger }$ and $c_{-1,i,s}^{\dagger }$ are electron creation
operators respectively to the right and left of the domain wall site $%
n_{0,i} $. (t-J model projection operator which excludes double occupancy is
implicitly assumed through out the paper).

\subsection{Kinetic {\em transverse} hole fluctuations}

In the absence of hoping interaction, $t$, the anti-phase stripe groundstate
(a) and the in-phase stripe groundstate (a') are degenerate. \ But, the
energy of the corresponding excited states (b) and (b') differ. As a result, 
{\em transverse hole hoping interaction removes the ground state degeneracy
in favor of anti-phase domain wall}\cite{Note2}.

\begin{figure}
\begin{center}
\leavevmode\epsfxsize=3.1in 
\epsfbox{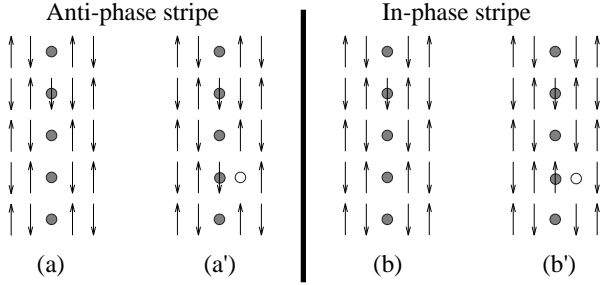} 
\end{center}
\caption{The domain wall sites are indicated by colored circles. The static 
ground state configuration of anti-phase and in-phase domain walls are 
depicted in figures (a) and (b) respectively. The corresponding excited states 
(a') and (b') result from transverse hoping dynamics of holes. Note the ''bad bond'' 
created by the hoping in (b'). The depicted domain wall groundstate segments 
have one electron per 5 sites on the domain wall 
(i.e., a filling fraction $\delta ^{e}=0.1$). Note the difference in the magnetic bonds 
of the domain wall electron between (a) and (b). }
\label{fig-1}
\end{figure}

To be general, we define $\varepsilon $ to be the energy difference between
an antiferromagnetic ''good bond'' and a ferromagnetic ''bad bond'' of two
neighboring spins. Obviously, $\varepsilon $ is proportional to $J$ (and $%
\varepsilon =J$ in the case of Ising model of the AFM spin domains). We
label by $E_{1}^{anti}$ and $E_{1}^{in}$ the bare excited states energy in
the case of an anti-phase and in-phase domain walls as depicted in figure ($%
1b$) and ($1b^{^{\prime }}$) respectively. Clearly, 
\begin{equation}
E_{1}^{in}-E_{1}^{anti}=\varepsilon  \label{epsilon-define}
\end{equation}
The Hamiltonian (\ref{H_site}) is thus given by the matrix 
\begin{equation}
H^{\alpha }=\left( 
\begin{array}{ccc}
0 & t & t \\ 
t & E_{1}^{\alpha } & 0 \\ 
t & 0 & E_{1}^{\alpha }
\end{array}
\right)
\end{equation}
($\alpha =$anti/in) which can be diagonalized exactly, resulting with the
ground state energy $E_{g}^{\alpha }$ 
\begin{eqnarray}
E_{g}^{\alpha } &=&\frac{1}{2}\left( E_{1}^{\alpha }-\sqrt{\left(
E_{1}^{\alpha }\right) ^{2}+8t^{2}}\right) \\
&\approx &\left\{ 
\begin{array}{ll}
\frac{1}{2}\left[ E_{1}^{\alpha }-\sqrt{8}t\left( 1+\frac{\left(
E_{1}^{\alpha }\right) ^{2}}{16t^{2}}\right) \right] & \text{ \ for }J\ll t
\\ 
-\frac{2t^{2}}{E_{1}} & \text{\ \ for }J\gg t
\end{array}
\right.  \nonumber
\end{eqnarray}
The difference in kinetic energy gain due to transverse hoping, between an
anti-phase and an in-phase domain wall configurations, is given by 
\begin{eqnarray}
\Delta E_{1}^{kin} &=&E_{g}^{anti}-E_{g}^{in}  \label{E_kinetic_trans} \\
&\approx &\left\{ 
\begin{array}{ll}
-\frac{\varepsilon }{2}+{\cal O}\left( \frac{\varepsilon ^{2}}{t}\right) & 
\text{ \ for }J\ll t \\ 
-\frac{t^{2}}{\varepsilon } & \text{\ \ for }J\gg t
\end{array}
\right.
\end{eqnarray}
Our result agrees with a previous large-d calculation (valid only in the
limit $J\gg t$)\cite{Large-d-Erica98}. But, in the experimental systems of
interest $\frac{J}{t}\approx \frac{1}{3}$, and thus they are better
approximated by calculations in the limit $J\ll t$ which will be assumed for
the rest of our discussion. From equation (\ref{E_kinetic_trans}) we
conclude that, {\em in a site-centered stripe geometry, the zero point
transverse kinetic fluctuation of the holes favor an antiferromagnetic
alignment of the neighboring spins}.

Note that the above result is in stark contrast with the conventional wisdom
that hole's zero-point kinetic fluctuations always favors ferromagnetic
alignment of their environment\cite{Zener-DoubleExchange,Aurbach-Hole-ferro}%
, which is based mostly on isolated hole models. It is a rather simple
demonstration of the difference between collective and single hole
properties in doped antiferromagnets.

\subsection{Competing magnetic interactions at electron filling $\protect%
\delta \neq 0$}

From the analysis of the above model, we conclude that the {\em transverse
kinetic fluctuations of holes are sufficient to induce an anti-phase domain
wall ground state in an empty domain wall }$\delta =0${\em \ (i.e., one hole
per site along the wall, }$n^{h}=1${\em ) }of stripes.

For a domain wall with electron filling fraction of $\delta $, $2\delta $ is
the number of electrons per site along the domain wall (in the large U
limit), and $n_{h}=\left( 1-2\delta \right) $ is the number of holes in the
lower Hubbard band. Each hole contributes a kinetic energy difference $%
\Delta E_{1}^{kin}$. Hence, at an arbitrary electron filling fraction $%
\delta $ of the domain wall, the average {\em transverse kinetic energy gain
per site} $\Delta E_{\perp }^{kin}\left( \delta \right) $ of an anti-phase
domain wall in comparison with an in-phase domain wall is

\begin{equation}
\Delta E_{\perp }^{kin}\left( \delta \right) =\frac{\Delta E_{1}^{kin}}{%
N_{site}}=-\frac{\varepsilon }{2}\left( 1-2\delta \right)
\label{E_hole_transverse}
\end{equation}
For each electron on the domain wall there is an energy difference $\Delta
E_{1}^{mag}=+\varepsilon $ in favor of an in-phase domain wall, due to spin
exchange interaction with the AFM environment. At electron filling $\delta $%
, the average magnetic energy difference (of an anti-phase domain wall in
comparison with an in-phase domain) per site along the domain wall is 
\begin{equation}
\Delta E_{\perp }^{mag}\left( \delta \right) =\frac{\Delta E_{1}^{mag}}{%
N_{site}}=+\varepsilon \left( 2\delta \right) .  \label{E_mag-static}
\end{equation}
{\em The competition of these transverse interaction alone would predict a
transition as a function of domain wall filling from anti-phase to in-phase
domain wall when} 
\begin{equation}
\Delta E_{\perp }\left( \delta \right) =\Delta E_{\perp }^{kin}+\Delta
E_{\perp }^{mag}>0
\end{equation}
{\em \ at } 
\begin{equation}
\delta \geq \frac{1}{6}  \label{delta_c transverse}
\end{equation}
(corresponding to $n^{h}\leq 2/3$, i.e., at a density of two holes per three
sites), in conflict with experimental observations of anti-phase stripes in $%
\left( LaNd\right) SrCuO$ materials with $\delta \approx 1/4$. This conflict
is resolved in the next section, where we account for effect of kinetic
dynamics {\em along} the domain wall.

\section{Implications of electron dynamics {\em along} the domain wall}

In this section we investigate the consequences of electron dynamics along
the domain wall for the competition between anti-phase and in-phase domain
wall configurations. As a first order approximation, we assume static spin
configuration of the antiferromagnetic domains on the left and right of the
domain wall. Thus, electrons moving along the domain wall are effectively
modeled as a one-dimensional electron gas (1DEG) in a static external
magnetic field. At this mean-field level, electrons moving in an anti-phase
domain wall experience a net zero external field on each site, while
electrons moving on an in-phase domain wall experience a staggered external
magnetic field of magnitude proportional to the spin interaction strength $J$%
. Hence, the effective domain wall 1DEG Hamiltonian has the form 
\begin{eqnarray}
H &=&H_{0}\left( B\right) +U\sum_{j}n_{j\uparrow }n_{j\downarrow }
\label{H_wall} \\
H_{0}\left( B\right) &=&t_{\parallel }\sum_{j\sigma }\left( c_{j\sigma
}^{\dagger }c_{j+1,\sigma }+h.c.\right) -\mu _{F}\sum_{j\sigma }c_{j\sigma
}^{\dagger }c_{j\sigma }  \label{H0(B)} \\
&&+2B\sum_{j\sigma }\sigma \left( -1\right) ^{j}c_{j\sigma }^{\dagger
}c_{j\sigma }  \nonumber
\end{eqnarray}
where $\sigma =\pm 1$ for spin $\uparrow ,\downarrow $ respectively, and 
\begin{equation}
B=\left\{ 
\begin{array}{cc}
0 & \text{ \ \ \ for anti-phase} \\ 
\approx \frac{J}{4} & \text{ for in-phase}
\end{array}
\right.
\end{equation}

When considering the competition of stripes versus droplets forms of local
phase separation, Nayak\&Wilzek\cite{Nayak-stripes} proposed that a
significant energy is gained by the increased mobility {\em along} an
anti-phase domain wall. But we note that the kinetic energy gain in an
in-phase domain wall is practically the same as in anti-phase domain wall
state. Therefore, it is essential to make a more careful analysis of the
electronic dynamics along the domain wall in both cases.

Below, we extract quantitative results in two limits: (a) For
non-interacting electrons moving along the domain wall, and (b) In the large 
$U\gg t$ limit for the interaction of electrons along the domain wall. In
both cases, the essential result is that kinetic fluctuations along the
domain wall weakens the magnetic interaction energy gain which favors an
in-phase domain wall environment, and thus extends the stability of
anti-phase domain wall configuration to higher electron filling fractions
(i.e., lower hole densities). In particular,{\em \ }we conclude that
dynamics fluctuations along the domain wall allow for the establishment of
anti-phase stripes with domain wall filling $\delta \approx \frac{1}{4}$\
(as in Copper-Oxides systems).

\subsection{Effect of motion along the domain wall for non-interacting
electrons at arbitrary filling}

We here model the dynamics along the hole rich domain wall by an effective
one dimensional lattice model (\ref{H_wall}) of non-interacting electrons ($%
U=0$). For an in-phase domain wall, the exchange coupling to the spins in
the AFM environment result with an effective external staggered magnetic
field on the electrons moving along the domain wall (see figure-1b). Our
model of kinetic motion along the domain wall included only band structure
effects due to a static spin configuration of the AFM environment, but no
dynamic scattering interactions (leading to finite resistivity)\cite{Note4}.

The anti-phase domain wall spectrum ($B=0$) is 
\begin{eqnarray}
E_{n}^{(anti)} &=&-2t_{\parallel }\cos \left( k_{n}a\right) -\mu
_{F}^{(anti)} \\
k_{n}a &=&\frac{2\pi }{N}n\text{ \ \ (}-\frac{N}{2}\leq n\leq \frac{N}{2}%
\text{)} \\
\mu _{F}^{(anti)} &=&-2t_{\parallel }\cos \left( \frac{2\pi }{N}n_{F}\right)
\end{eqnarray}
where $N$ is the number of sites.

For an in-phase domain wall ($B\neq 0$), the staggered field doubles the
unit cell. The non-interacting Hamiltonian (\ref{H0(B)}) $H_{0}\left( B\neq
0\right) $ is diagonalized by the appropriate Bloch states; 
\begin{eqnarray}
\psi _{k,\sigma }^{\dagger } &=&\sqrt{\frac{2}{N}}\sum_{j=1}^{N/2}e^{+ik%
\left( 2ja\right) }W_{j,\sigma }\left( k\right)
\label{Wj(k) staggered basis} \\
W_{j,\sigma }\left( k\right) &=&\frac{1}{\sqrt{1+\left| f_{k,\sigma }\right|
^{2}}}\left[ c_{2j,\sigma }^{\dagger }+f_{k,\sigma }e^{-ika}c_{2j-1,\sigma
}^{\dagger }\right]  \nonumber \\
f_{k,\sigma } &=&\frac{-\sigma \left( \frac{B}{t}\right) \mp \sqrt{\left( 
\frac{B}{t}\right) ^{2}+\cos ^{2}\left( k\right) }}{\cos \left( k\right) } 
\nonumber
\end{eqnarray}
The resulting energy spectrum for the in-phase domain wall is 
\begin{eqnarray}
&&E_{n}^{(in)}=\pm 2t_{\parallel }\sqrt{\left( \frac{B}{t_{\parallel }}%
\right) ^{2}+\cos ^{2}\left( \frac{2\pi }{N}n\right) }-\mu _{F}^{(in)}
\label{En} \\
\text{ \ } &&\text{(}-\frac{N}{4}\leq n\leq \frac{N}{4}\text{)}
\end{eqnarray}
\begin{equation}
\delta =\frac{2n_{F}}{N}
\end{equation}
In this context we comment that for a Hubbard model in an external staggered
magnetic field, unlike the case of staggered charge potential/interaction,
there is no charge gap opening at 1/4 filling\cite{Zachar-unpublished}. For
simplicity, the rest of the formulas are written for the case where the
odd-sites are with the magnetic field anti-parallel to the spin. Moreover,
we remind the reader that we assume $B\sim J\ll t_{\parallel }$. Well below
half filling, (i.e., $t_{\parallel }\cos \left( k\right) >0$), 
\begin{eqnarray}
f_{k,\uparrow } &=&\frac{-B+\sqrt{B^{2}+t_{\parallel }{}^{2}\cos ^{2}\left(
k\right) }}{t_{\parallel }\cos \left( k\right) }  \nonumber \\
&\approx &1-\frac{B}{t_{\parallel }\cos \left( k\right) }+{\cal O}\left( 
\frac{B}{t_{\parallel }\cos \left( k\right) }\right) ^{2}
\end{eqnarray}

The gain in magnetic energy per $k$-state is due to the difference between
the occupation probability of odd and even sites for a given electron spin,
(this is the main difference between 1DEG in an in-phase vs. anti-phase
domain wall), 
\begin{eqnarray}
\left| \left\langle c_{2j,\sigma }^{\dagger }c_{2j,\sigma }\right\rangle
_{k}\right| ^{2}- &&\left| \left\langle c_{2j-1,\sigma }^{\dagger
}c_{2j-1,\sigma }\right\rangle _{k}\right| ^{2} \\
&\approx &\sum_{j=1}^{N/2}\frac{2}{N}\frac{1}{1+\left| f_{k,\sigma }\right|
^{2}}\left( 2\frac{B}{t_{\parallel }\cos \left( k\right) }\right)   \nonumber
\\
&\approx &\left( \frac{B}{t_{\parallel }\cos \left( k\right) }\right) +{\cal %
O}\left( \frac{B}{t_{\parallel }\cos \left( k\right) }\right) ^{2}  \nonumber
\end{eqnarray}
Therefore, for filling up to $k_{F}$, the magnetic interaction energy gain
for an in-phase domain wall in comparison with an anti-phase domain wall is
given approximately by, 
\begin{eqnarray}
\Delta E^{mag}\left( \delta \right)  &\approx &2B\left[ 2\int_{-\pi \delta
}^{\pi \delta }\frac{B}{t_{\parallel }\cos \left( k\right) }dk\right]  \\
&=&2B\left[ 2\left( \frac{B}{t_{\parallel }}\right) \ln \left( \frac{1+\sin
\left( \pi \delta \right) }{1-\sin \left( \pi \delta \right) }\right) \right]
\\
&\approx &\frac{J}{4}\left[ \frac{J}{t_{\parallel }}\ln \left( \frac{1+\sin
\left( \pi \delta \right) }{1-\sin \left( \pi \delta \right) }\right) \right]
\label{E_mag_free}
\end{eqnarray}
(Note that the limit $t_{\parallel }\rightarrow 0$ of the previous section
cannot be recovered since we used approximations based on $t_{\parallel }\gg
B\approx J/4$).

We are now in a position to give a first analytical estimate of the critical
transition point filling $\delta _{c}$ determined given by (using (\ref
{E_hole_transverse}) and (\ref{E_mag_free})) 
\begin{eqnarray}
0 &=&\Delta E_{\perp }\left( \delta _{c}\right) \approx E_{\perp
}^{kin}+\Delta E^{mag} \\
&\approx &-\frac{J}{2}\left( 1-2\delta _{c}\right) +\frac{J}{4}\left[ \frac{J%
}{t}\ln \left( \frac{1+\sin \left( \pi \delta _{c}\right) }{1-\sin \left(
\pi \delta _{c}\right) }\right) \right]
\end{eqnarray}
The geometrical solution for $\frac{J}{t}=\frac{1}{3},\frac{1}{4},\frac{1}{6}%
,\frac{1}{8},\frac{1}{10}$ is plotted in figure-2

\begin{figure}
\begin{center}
\leavevmode\epsfxsize=3.1in 
\epsfbox{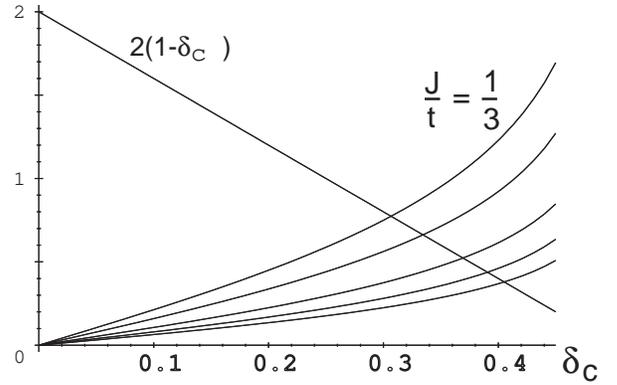} 
\end{center}
\caption{geometrical solution for $\frac{J}{t}=\frac{1}{3},\frac{1}{4},\frac{1}{6},
\frac{1}{8},\frac{1}{10}$ }
\label{fig-2}
\end{figure}

The main qualitative result is that {\em electron hoping fluctuations along
a partially filled domain wall inhibit the effective magnetic interaction
energy across the domain wall}. As a result, an anti-phase domain wall
groundstate configuration can be obtained beyond $\delta =\frac{1}{6}$,
depending on the value of $J/t$. For $\frac{1}{3}\leq \frac{J}{t}\leq \frac{1%
}{10}$ we find $0.3<\delta _{c}<0.4$. Note that the critical filling
fraction $\delta _{c}$ is not very sensitive to the value of $J/t$. Most
importantly, it is above the value $\delta \approx 0.25$ of anti-phase
stripes in $\left( LaNd\right) SrCuO$ systems\cite{Tranquada-reviews}.

\subsection{Large $U\gg t$ model, and a second estimate of the critical
domain wall filling $\protect\delta _{c}$}

In this subsection we examine the competition between anti-phase and
in-phase domain wall configurations for strongly interacting electrons, $%
U\gg t$, along the domain wall at arbitrary filling fraction.

The exact ground-state energy of the Hubbard model (\ref{H_wall}) is well
known for the case $B=0$ from Bethe ansatz solution. Unfortunately, we do
not know of an established solution for the groundstate energy of a one
dimensional Hubbard model in a staggered magnetic field $B\neq 0$. In
particular, when $B=0$ the Hubbard interaction in momentum space takes the
form $Un_{k\uparrow }n_{k\downarrow }$, and the $U=\infty $ limit is
effectively captured by imposing a restriction of one electron per $k-$%
state. But such a simple representation does not exist in terms of the $%
W_{j,\sigma }\left( k\right) $ basis (\ref{Wj(k) staggered basis}) in a
staggered external magnetic field. Therefore, we here develope a way to
approximate the two competing wall configuration energies at the same level
of approximation.

We wish to tress that the qualitative effect of longitudinal hoping
fluctuations - to inhibit the effective magnetic interaction energy across
the domain wall - was already established in the preceding subsection. The
purpose of the current subsection is to investigate the quantitative effect
of including large Hubbard interactions between the electrons. Though we
indeed resort to approximations at several stages of our model and
calculations, we do capture the substantial effect. Once the core physics
established, the degree to which our final numerical values are exact can be
checked (and improved) in numerical simulations on finite systems.

Our main result is that the {\em large Hubbard interaction}, $U$, {\em %
delimitate the critical filling fraction to be in the narrow range} 
\begin{equation}
0.28<\delta _{c}<0.30  \label{delta_c large-U}
\end{equation}
{\em for any value of }$\frac{J}{t}<\frac{1}{3}$.

Below, we explain our modeling approach and calculation. We first apply the
model to examine the simplest case of 1/4 filled domain wall, and show that
always an anti-phase is obtained. Then, we apply the model to re-estimate
the critical filling fraction for a transition to an in-phase domain wall,
and thus establish the result noted above in equation (\ref{delta_c large-U}%
).

\subsubsection{Model of characteristic unit cells}

The unit cell of a stripe domain wall comprise of two lines as depicted in
figure-3. Our modeling idea in this subsection is to{\em \ solve exactly the
Hamiltonian of isolated prototypical unit cells, and then approximate the
full stripe as assembled of a collection of such unit cells}. Such an
approximation is building on our findings in the previous subsection, where
we argued that it is not the global longitudinal mobility, but only the
local kinetic fluctuations (in the occupation of odd/even numbered sites
along the wall) which distinguish anti-phase and in-phase domain walls.

As shown in figure-3, in the $U=\infty $ limit, there are only three
possible electron occupation numbers of a unit cell: one, two or zero
electrons on the domain wall sites. In our model approximation, we ignore
the hopping from one unit cell to another. As we shall later explain, for
filling fractions $\delta \geq \frac{1}{4}$ it will suffice to consider only
unit cells with one and two electrons on the domain wall sites (i.e., as in
figure-3(a1,a2 and b1,b2)).

\begin{figure}
\begin{center}
\leavevmode\epsfxsize=3.1in 
\epsfbox{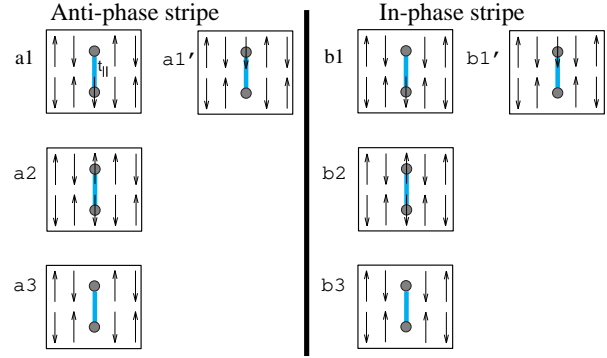} 
\end{center}
\caption{Characteristic unit cells of a stripe domain wall. Any domain wall 
groundstate can be constructed by properly assembling these unit cells. 
For domain wall groundstate with filling fraction $\delta ^{e}>1/4$, 
unit cells (a3) and (b3) can be neglected. }
\label{fig-3}
\end{figure}

{\em We are interested only in the effect of hopping dynamics }$t_{\parallel
}${\em \ of electron along the domain wall. }Therefore we examine first the%
{\em \ unit cells with one hole and one electron on the domain wall sites}.
There's spin exchange interactions of strength $J$ between nearest-neighbor
electrons. Yet, to zeroth level of approximation, the spin configuration is
assumed to be static (apart from $t_{\parallel }$ electron hoping). Thus, we
consider only the electron configurations of an anti-phase (a1,a1') and
in-phase (b1,b1') domain wall unit cells.

Consider the case of unit cells with a single hole and spin-$\downarrow $
electron hopping between the domain wall sites. We observe that while the
anti-phase domain wall states (a1,a1') are degenerate, the in-phase domain
wall states (b1 and b1') have an energy difference of magnitude $%
2\varepsilon $ (where $\varepsilon \sim J$ is the energy difference between
parallel and anti-parallel neighboring spin states). Thus, the Hamiltonian
of a single electron in an anti-phase domain wall unit cell is 
\begin{equation}
H_{\text{hole-cell}}^{anti}=\left( 
\begin{array}{cc}
0 & -t_{\parallel } \\ 
-t_{\parallel } & 0
\end{array}
\right)  \label{H_cell-anti}
\end{equation}
with groundstate energy $-t$, while for an in-phase domain wall 
\begin{equation}
H_{\text{hole-cell}}^{in}=\left( 
\begin{array}{cc}
-\varepsilon & -t_{\parallel } \\ 
-t_{\parallel } & +\varepsilon
\end{array}
\right)  \label{H_cell-in}
\end{equation}
with groundstate energy $-\sqrt{\left( \varepsilon ^{2}+t_{\parallel
}^{2}\right) }\approx -\left[ t_{\parallel }+\frac{1}{2}\left( \frac{%
\varepsilon }{t_{\parallel }}\right) \varepsilon \right] $. (Note that the
magnetic interaction energy is already fully accounted for in the
eigenvalues, so it shouldn't be counted again). Thus we find that, for a
stripe unit cell with one hole and one electron on the domain wall sites,
the ground state magnetic interaction energy difference between an
anti-phase and in-phase domain wall is 
\begin{eqnarray}
\Delta E_{\text{hole-cell}}^{mag} &=&E_{\text{hole-cell}}^{anti}-E_{\text{%
hole-cell}}^{in}  \label{E_mag-hole-cell} \\
&=&-t_{\parallel }+\sqrt{\left( \varepsilon ^{2}+t_{\parallel }^{2}\right) }
\nonumber \\
&\approx &\left\{ 
\begin{array}{ll}
+\frac{1}{2}\left( \frac{\varepsilon }{t_{\parallel }}\right) \varepsilon & 
\text{\ \ for }t_{\parallel }\gg J \\ 
+\varepsilon & \text{\ \ for }t_{\parallel }\rightarrow 0
\end{array}
\right.  \nonumber
\end{eqnarray}
The above needs to be divided by $2$ in order to get the energy difference
per site, since the unit cell has two sites along the domain wall. In the
static limit $t_{\parallel }\rightarrow 0$, we recover equation (\ref
{E_mag-static}) of section-III, $\frac{\Delta E^{mag}\left( \delta
=1/4\right) }{N}\approx +\frac{\varepsilon }{2}$. But for the rest of the
paper, we focus on the experimentally more relevant limit $t_{\parallel }\gg
J$.

We now proceed to demonstrate first the essence of our approach for the
simplest case of $\delta =\frac{1}{4}$ filling.

\subsubsection{The special case of 1/4 filling}

In section-III, we have shown that dynamics of only {\em transverse}
interactions (hole hopping and Heisenberg interaction) will not suffice to
stabilize an anti-phase domain wall for $\delta >1/6$. Therefore, by
investigating the case of $\delta =1/4$ we can already answer the question
of principle about the effect of electron dynamics {\em along} a stripe
domain wall. Moreover, it is a case of particular interest for stripes in
HTc, which are found with a filling fraction very near $\delta =1/4$.

In the limit of large on-site interaction $U\gg t$, at $\delta =1/4$ there
is effectively one electron for every two sites along the wall. Moreover, to
mimic the expected effect of coulomb interactions, it makes sense to
supplement the model with a repulsive interaction also between near-neighbor
holes (i.e., a $t-U-V$ model along the domain wall)\cite{Leonid00-StripeWall}%
. Thus, in the absence of kinetic hoping dynamics along the domain wall, the
groundstate of the one dimensional electron gas is a periodic charge density
wave (CDW) with one electron per every second site. Even with the inclusion
of kinetic fluctuations, the same CDW still dominate the electron
correlations at low temperatures. As a result, we argue that the occurrence
of unit cells with an occupation of zero or two electrons on the domain wall
sites (as in figure-3(a2,a3)) are rare events and thus their contribution
may be safely neglected.

Therefore, a two-line stripes segment with one hole and one electron on the
domain wall sites can be regarded as the characteristic unit cell along a
1/4 filled domain wall (as in figure-3(a1,b1)). By solving such unit cell
model exactly, we are able to account for the effect of local kinetic
fluctuations along the domain wall. Moreover, since we are interested only
in the difference between anti-phase and in-phase wall, and since the net
magnetic field averages to zero along both anti-phase and in-phase wall, it
is intuitively expected that {\em the significant difference is not in the
net electron mobility but mostly in the local kinetic fluctuations} (Though
magnetic scattering from AFM environment, which we neglect, may also differ).

Let us recall first the result of section-III. In the absence of hopping
along the wall, the average transverse kinetic energy per site of an
anti-phase domain wall in comparison with an in-phase domain wall was (using
equation (\ref{E_hole_transverse}))

\[
\Delta E_{\perp }^{kin}\left( \delta =\frac{1}{4}\right) =\frac{E_{\perp
}^{kin}}{N}=-\frac{\varepsilon }{2}\left( 1-2\delta \right) =-\frac{%
\varepsilon }{4} 
\]
and the competing transverse magnetic energy due to Heisenberg interaction
with the AFM environment was (using equation (\ref{E_mag-static})) 
\[
\Delta E_{\perp }^{mag}\left( \delta =\frac{1}{4}\right) =\frac{E_{\perp
}^{mag}}{N}=+\varepsilon \left( 2\delta \right) =+\frac{\varepsilon }{2}. 
\]
Hence, it was the in-phase domain wall configuration which had the lowest
energy ground state. Now, with the inclusion of the effect of the kinetic
fluctuations along the domain wall, the average magnetic interaction energy
is modified and (using (\ref{E_mag-hole-cell}) instead of (\ref{E_mag-static}%
)) the average energy difference per site is 
\begin{eqnarray}
\frac{\Delta E}{N} &=&\Delta E_{\perp }^{kin}+\frac{1}{2}\Delta E_{\text{%
hole-cell}}^{mag}  \label{1/4 result} \\
\frac{\Delta E\left( \delta =1/4\right) }{N} &\approx &-\frac{\varepsilon }{4%
}+\left( \frac{\varepsilon }{t_{\parallel }}\right) \frac{\varepsilon }{4}<0
\end{eqnarray}
where $N$ is the number of sites along the domain wall, (The factor 1/2 in (%
\ref{1/4 result}) is due to having two domain wall sites per unit cell).
Therefore, we find that due to the kinetic fluctuations along the domain
wall, the magnetic interaction energy gain of an in-phase domain wall
configuration is reduced enough to make the anti-phase domain wall more
favorable at 1/4 filling for any value of $J\sim \varepsilon <t_{\parallel }$%
.

\subsubsection{Transition at domain wall filling $\protect\delta _{c}$}

We now attempt to determine the contribution of the effect described above
at arbitrary electron filling $2\delta >\frac{1}{2}$, and thus estimate the
transition point.

As argued before, the 1DEG correlations on the domain wall are dominated by
a CDW correlations where, at electron filling $2\delta >1/2$, the occurrence
of two near-neighbor holes is statistically negligible compared with other
configurations in the ground state. Therefore, the two-sites unit cells
along the domain wall are basically only of two kinds: (a) There are $\left(
1-2\delta \right) $ unit cells containing one hole, which are the type
analyzed in the previous subsection. (b) There are $\frac{1}{2}-\left(
1-2\delta \right) =\left( 2\delta -\frac{1}{2}\right) $ unit cells
containing no holes.

To estimate the average energy contribution per site, we can use the result
of the previous subsection for the hole cells, only with the added factor $%
\left( 1-2\delta \right) $;

\begin{eqnarray}
\Delta E^{a}\left( \delta \right) &=&\Delta E_{\perp }^{kin}+\frac{1}{2}%
\Delta E_{\text{hole-cell}}^{mag} \\
&=&-\frac{\varepsilon }{4}\left[ 1-\left( \frac{\varepsilon }{t_{\parallel }}%
\right) \right] \left( 1-2\delta \right)
\end{eqnarray}
and add the contribution of magnetic energy difference per site due to unit
cells with no holes, 
\begin{equation}
\Delta E^{b}\left( \delta \right) =\frac{1}{2}E_{\text{2e-cell}}^{mag}=+%
\frac{1}{2}\left( 2\delta -\frac{1}{2}\right) 2\varepsilon .
\end{equation}
The preferred groundstate configuration is determined by 
\begin{eqnarray}
\frac{\Delta E\left( \delta \right) }{N} &=&\Delta E^{a}\left( \delta
\right) +\Delta E^{b}\left( \delta \right)  \label{E_anti - E_in} \\
&=&-\frac{\varepsilon }{4}\left[ 1-\left( \frac{\varepsilon }{t_{\parallel }}%
\right) \right] \left( 1-2\delta \right) +\left( 2\delta -\frac{1}{2}\right)
\varepsilon .
\end{eqnarray}
A transition from anti-phase to in-phase domain wall occurs when 
\[
\Delta E\left( \delta \right) >0 
\]
Therefore, the critical filling fraction $\delta _{c}$ is given by{\em \ } 
\begin{eqnarray}
0 &=&-\frac{\varepsilon }{2}\left[ 1-\left( \frac{\varepsilon }{t_{\parallel
}}\right) \right] \left( 1-2\delta _{c}\right) +2\left( 2\delta _{c}-\frac{1%
}{2}\right) \varepsilon \\
\delta _{c} &=&\frac{\frac{3}{2}-\frac{1}{2}\left( \frac{\varepsilon }{%
t_{\parallel }}\right) }{5-\left( \frac{\varepsilon }{t_{\parallel }}\right) 
}
\end{eqnarray}
The predicted $\delta _{c}$ for several values of $\frac{\varepsilon }{t}$
is given below, 
\begin{eqnarray*}
\delta _{c}\left( \frac{\varepsilon }{t}=\frac{1}{3}\right)
&=&0.\,\allowbreak 285\,71 \\
\delta _{c}\left( \frac{\varepsilon }{t}=\frac{1}{6}\right)
&=&0.\,\allowbreak 293\,1 \\
\delta _{c}\left( \frac{\varepsilon }{t}\rightarrow 0\right) &\lessapprox
&0.\,\allowbreak 3
\end{eqnarray*}
{\em We find that }$0.28<\delta _{c}<0.3${\em , (for }$\frac{J}{t}<\frac{1}{3%
}${\em ), and is only weakly sensitive to the value of }$\frac{J}{t}$.

\section{Connection with a Landau theory of stripes order}

The properties of a general Ginzburg-Landau free energy (\ref
{Stripes-FreeEnergy}) of stripes were previously investigated in\cite
{StripesLandau,Leonid99-StripesLandau}. Specifically, an ordered stripe
phase is a unidirectional density wave which consists of coupled
spin-density wave (SDW) and charge-density wave (CDW) order parameters. 
\begin{eqnarray}
{\cal F}_{q,k} &=&r_{1{\rm s}}\,\left| {\bf S}_{\vec{q}}\right| ^{2}+r_{2%
{\rm s}}\,\left| {\bf S}_{\vec{q}+\vec{k}}\right| ^{2}+r_{{\rm c}}\,\left|
\rho _{\vec{k}}\right| ^{2}  \label{Stripes-FreeEnergy} \\
&&+\gamma \left[ {\bf S}_{\vec{q}}^{\ast }\,{\bf S}_{\vec{q}+\vec{k}}\,\rho
_{\vec{k}}^{\ast }+{\rm c.c.}\right] +\ldots ,  \nonumber
\end{eqnarray}
where $\rho _{\vec{k}}$ is the complex-valued CDW amplitude with wave vector 
$\vec{k}$, $\rho _{\vec{k}}^{\ast }\equiv \rho _{-\vec{k}}$, and similarly $%
{\bf S}_{\vec{q}}$ is a complex vector amplitude of the SDW. The quartic
(and higher order) terms required for stability are omitted.

Zachar {\sl et al}.\cite{StripesLandau} have considered the phase transition
between a stripe phase and a high-temperature disordered state, as involving
only a{\it \ single SDW wave vector (i.e., }${\bf S}_{\vec{q}}^{\ast }\,=%
{\bf S}_{\vec{q}+\vec{k}}${\it )}. The existence of the cubic $\gamma $
term, coupling these two order parameters in the Landau free energy, drives
the period of the SDW to be twice that of the CDW, and the absence of any
net AFM ordering is equivalent to the statement that the stripes are
topological.

In contrast, it was shown by Pryadko {\sl et al}.\cite
{Leonid99-StripesLandau}, that the same sort of cubic term in a Landau
theory of the transition from a homogeneous ordered antiferromagnetic phase
to a stripe ordered phase produces a state in which the antiferromagnetic
magnetization {\em does not\/} change its sign between the domains, {\it %
i.e.\/}\ the stripes are non-topological.

When investigating the transition {}from a well-developed antiferromagnet
with a modulation vector $\vec{\pi}=(\pi ,\pi )$ to an incommensurate
modulated phase, we must account for both the original AFM order parameter $%
{\bf S}_{\vec{\pi}}$ (which cannot be assumed small), and the spin density
wave ${\bf S}_{\vec{\pi}+\vec{k}}$. The most relevant terms in the Landau
expansion of the effective free energy are then 
\begin{eqnarray}
{\cal F} &=&r_{1{\rm s}}\,\left| {\bf S}_{\vec{\pi}}\right| ^{2}+r_{2{\rm s}%
}\,\left| {\bf S}_{\vec{\pi}+\vec{k}}\right| ^{2}+r_{{\rm c}}\,\left| \rho _{%
\vec{k}}\right| ^{2} \\
&&+\gamma \left[ {\bf S}_{\vec{\pi}}^{\ast }\,{\bf S}_{\vec{\pi}+\vec{k}%
}\,\rho _{\vec{k}}^{\ast }+{\rm c.c.}\right] +\ldots ,  \nonumber
\end{eqnarray}
where a finite $r_{1{\rm s}}<0$ is assumed as given. This expression implies
that an instability in either spin or charge sector generates both spin- and
charge-density waves at the wave vectors $\vec{q}=\vec{\pi}+\vec{k}$ and $%
\vec{k}$, respectively. Near the transition the magnitude of the
incommensurate peak is necessarily much smaller then the commensurate AFM
modulation $|{\bf S}_{\vec{\pi}+\vec{k}}|\ll |{\bf S}_{\vec{\pi}}|$. It is
easy to see that this corresponds to {\em in-phase\/} domain walls; The
derived relationship between $\vec{q}$ and $\vec{k}$ implies that the
periods of spin and charge modulation are equal.

Our analysis in this paper supply microscopic insight to the Landau theory
results. First, we find that the possibility of both topological and
non-topological stripe phases can indeed be realized within a single
microscopic model (as anti-phase and in-phase domain wall groundstates
respectively).

Second, we find that non-topological stripes (corresponding to in-phase
domain wall state) are indeed established due to enhanced antiferromagnetic
interactions (as first speculated by Castro-Neto\cite{NetoHone-Unpub}). Yet,
thus far the connection is more heuristic than rigorous.

Furthermore, after analyzing the effect higher order derivative terms in a
gradient expansion of the Ginzburg-Landau free energy, Pryadko {\sl et al}.
argue that\cite{Leonid99-StripesLandau}: When there is no frustration,
topological stripes are not established in the ground state. However they
speculate that frustration, such as competing first and second neighbor
interactions, or opposite sign terms in the gradient expansion of the
Ginzburg-Landau model ({\it i.e.\/}\ below a Lifshitz point), can stabilize
topological collinear domain walls. In other words, {\em topological stripes
are a consequence of physics on short length scale, and they do not appear
in a theory that considers only long-distance or low-energy physics}.

Pryadko {\sl et al}. speculate that the missing frustrating ingredient may
be due to inter-stripe interactions such as long range coulomb interactions
which are expected to exist between the charged domain walls. Our analysis
in this paper demonstrate that such interactions are not required. We have
shown that local single stripe dynamic interactions, (albeit involving
finite strength competing interactions, i.e., expectantly beyond the reach
of perturbative gradient expansion) are sufficient to realize both
alternative possible topological magnetic order across the charge line.

\section{Estimate of spin-wave velocity}

In this section, we use our microscopic domain wall model to estimate
spin-wave velocity reduction 
\begin{equation}
\alpha _{v}=\frac{v_{\perp }}{v_{0}},
\end{equation}
where $v_{\perp }$ is the spin-wave velocity perpendicular to the stripes
(i.e., along the modulation direction), and $v_{0}$ is the spin-wave
velocity in the undoped parent antiferromagnetic material. Equivalently, it
is the predicted spin-wave velocity anisotropy in stripes (where $v_{0}$ is
the same as the velocity along the stripes). Castro-Neto \& Hone\cite
{Neto-Hone} were the first to point out that within the stripes model, one
of the main effects of the hole rich lines would be to weaken the effective
exchange interaction between spins on either side across the domain wall.
Thus, a cardinal parameter in the model is the effective weakening factor of
the magnetic interaction $J_{W}$ across the stripes domain wall as compared
with the Heisenberg interaction $J$ within the hole free AFM regions (and
hence along the stripes). 
\begin{equation}
\alpha _{J}=\frac{J_{W}}{J}  \label{Jw/J}
\end{equation}
The {\em key new ingredient is our analytical estimate of }$\alpha _{J}$.
Consequently, while previous model estimates always involved fitting of free
parameters, we are able to estimate analytically the spin-wave velocity
anisotropy (without any free parameters).

Neto\&Hone\cite{Neto-Hone} proposed an effective anisotropic Heisenberg
model for describing the low temperature spin dynamics in a striped Cu-O
plane. Within the anisotropic Heisenberg model \cite{Neto-Hone}, they
derived the spin-wave velocity anisotropy 
\begin{equation}
v_{\perp }\approx \sqrt{\frac{\alpha _{J}}{2}}v_{0.}
\label{c_perp Neto-Hone}
\end{equation}
i.e., $\alpha _{v}=\sqrt{\frac{\alpha _{J}}{2}}$. By fitting other
consequences of their model to experimental results, Neto\&Hone extracted
the value 
\[
\alpha _{J}\approx 0.01. 
\]
The fitted value of $\alpha _{J}$, together with (\ref{c_perp Neto-Hone}) $%
\hbar v_{\perp }\approx 50meV-$\AA . As discussed by Tranquada {\sl et al}. 
\cite{Tranquada99-(LaNd)SrCuO-glass}, this value is much too small to be
compatible with inelastic neutron scattering experiments. It is further
incompatible with the range of energies over which stripes
incommensurability is observed.

Aiming specifically to model the interaction between narrow stripe AFM
regions, Tworzydlo {\sl et al}.\cite{Tworzydlo-Ladders-anisotropy.} made
predictions based on models of coupled narrow spin ladders (e.g., 2-leg or
3-leg ladders). The resulting expressions for the spin wave velocity
anisotropy were 
\begin{equation}
\alpha _{v}=\left\{ 
\begin{array}{cc}
\sqrt{\frac{2\alpha _{J}}{1+\alpha _{J}}} & \text{\ for 2-leg ladder} \\ 
\sqrt{\frac{3\alpha _{J}}{1+2\alpha _{J}}} & \text{\ for 3-leg ladder}
\end{array}
\right.  \label{Ladders-Anisotropy}
\end{equation}

In terms of 2-leg ladder model, Tranquada {\sl et al}. proclaimed to fit the
experiments well with a value of 
\[
\alpha _{J}\approx 0.35. 
\]
giving 
\[
\alpha _{v}\approx 0.\,\allowbreak 72 
\]
In principle, the ladder model results are quite sensitive to the width of
the stripes (which affects the ladder spin gap magnitude). Yet accidentally,
the above value of $\alpha _{J}\approx 0.35$ would give $\alpha _{v}\approx
0.\,\allowbreak 78$ for 3-leg ladder, i.e., practically the same as for
2-leg.

Our microscopic model is based solely on the local physics in the
neighborhood of the hole line. As such, it is insensitive to details of the
width of the AFM spin domains. We argue that, by definition, the energy
difference between alternative spin configurations across the domain wall is
a measure of the effective spin interaction across the domain wall in any
appropriate low energy theory. Therefore, using our microscopic model
calculation, we suggest that 
\begin{equation}
J_{W}=-\Delta E\left( \delta \right) =E^{in}-E^{anti}
\end{equation}
were $\Delta E\left( \delta \right) $ is given in equation (\ref{E_anti -
E_in}) 
\[
\Delta E\left( \delta \right) \approx -\frac{\varepsilon }{2}\left[ 1-\left( 
\frac{\varepsilon }{t}\right) \right] \left( 1-2\delta \right) +2\left(
2\delta -\frac{1}{2}\right) \varepsilon 
\]
(where as discussed previously, $\varepsilon \approx J$). Thus we can
calculate $\alpha _{J}$ and then, following the anisotropic Heisenberg model
reasoning of Castro-Neto \& Hone, use (\ref{c_perp Neto-Hone}) to calculate
the spin-wave velocity anisotropy.

For the particular case of stripes observed in LaNdSrCuO\cite
{Tranquada-reviews} $\delta \approx 1/4$, we obtain

\begin{equation}
J_{W}\approx \Delta E\left( \delta =\frac{1}{4}\right) \approx \frac{J}{4}%
\left[ 1-\left( \frac{J}{t}\right) \right]
\end{equation}
Substituting $\frac{1}{3}>\frac{J}{t}>\frac{1}{10}$, we obtain 
\begin{equation}
0.\,\allowbreak 17\,<\alpha _{J}\left( \delta \approx 1/4\right)
<0.\,\allowbreak 23
\end{equation}
Importantly, note that $\alpha _{J}\left( \delta \approx 1/4\right) $ is
rather insensitive to the variation of $\frac{J}{t}<\frac{1}{3}$. (where $%
\delta \approx 1/4$ is the supposed electron\ filling fraction of the stripe
domain walls in Copper-Oxides).

Using equation (\ref{c_perp Neto-Hone}) thus give an estimate of 
\begin{equation}
0.29<\alpha _{v}<0.\,\allowbreak 34
\end{equation}
or using equation (\ref{Ladders-Anisotropy}) give an estimate of 
\begin{equation}
\begin{array}{cc}
0.67<\alpha _{v}<0.71 & \text{\ for 2-leg ladder domains} \\ 
0.74<\alpha _{v}<0.78 & \text{\ for 3-leg ladder domains}
\end{array}
\end{equation}
which fit remarkably well with the experimentally deduced spin wave velocity
inhibition $0.60<\alpha _{v}<0.72$ in doped Copper-Oxides compared with the
undoped parent AFM material\cite{Experiment-SpinWaveVelocity}. Note that our
estimate is obtained by directly substituting the experimental value of $%
v_{0}$ into the analytical results for $\alpha _{J}$ (using our model
equation (\ref{E_anti - E_in})) and for $\alpha _{v}$ (from the ladder
models \cite{Tworzydlo-Ladders-anisotropy.} or the anisotropic Heisenberg
model\cite{Neto-Hone}), without any fitting of free parameters.

\bigskip

{\bf Acknowledgments}: I thank Natan Andrei, Thierry Giamarchi, Baruch
Horovitz, Efstratios Manousakis, John Tranquada, and particularly Steve
Kivelson for fruitful discussions. This work was partly supported by the
TMR\#ERB4001GT97294 fellowship.

\end{document}